# Learning Diagnosis of COVID-19 from a Single Radiological Image

Pengyi Zhang, Yunxin Zhong, Xiaoying Tang, Yunlin Deng, Xiaoqiong Li[*]

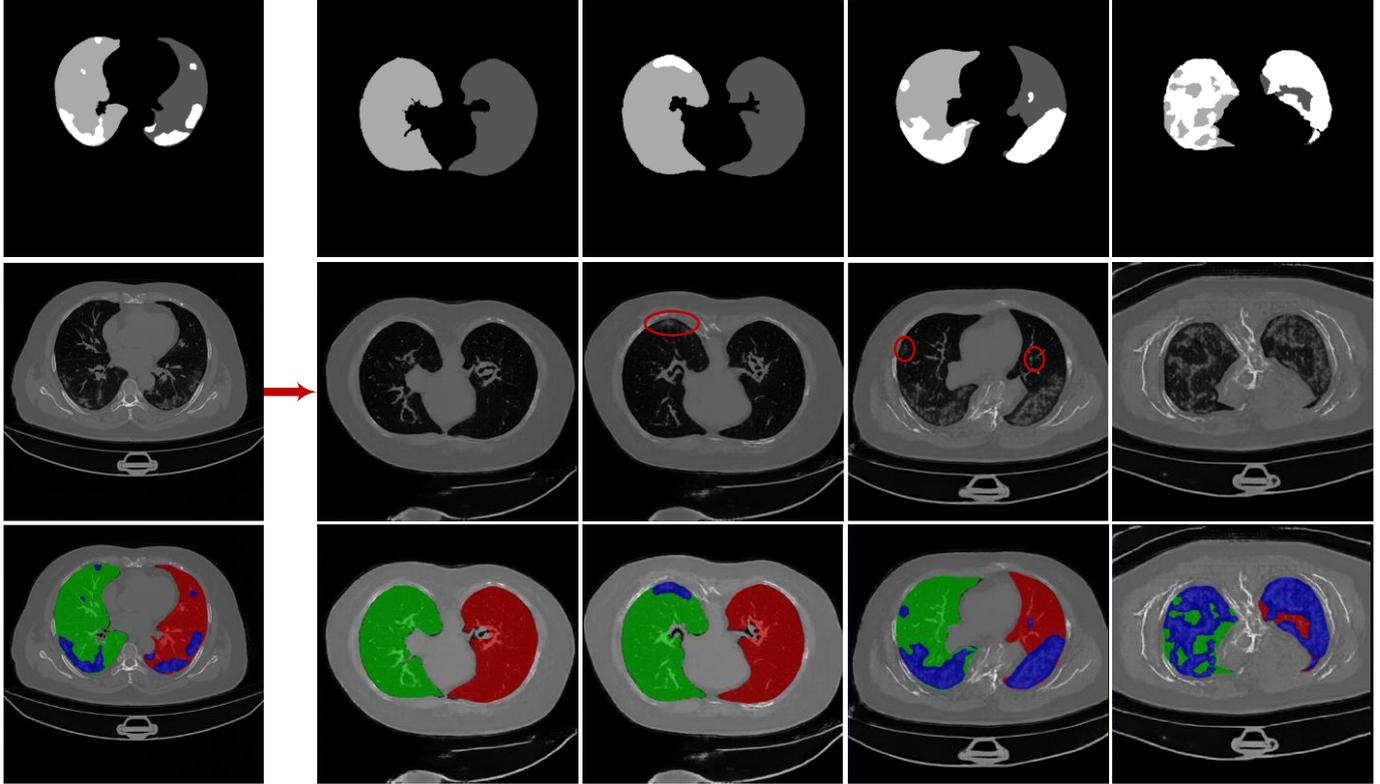

**Figure 1**. High-resolution (512×512) radiological images with COVID-19 infection synthesized by proposed CoSinGAN. CoSinGAN is a novel conditional generative model trained on a single chest CT slice with pixel-level annotation mask of the lung and COVID-19 infection. Our model is able to capture the conditional distribution of visual finds of the lung and COVID-19 infection accurately. We explore the feasibility of learning deep models for COVID-19 diagnosis from a single radiological image by resorting to synthesizing diverse radiological images with pixel-level annotations of COVID-19 infection. Both classification and segmentation networks trained on these synthesized radiological images achieve notable detection accuracy of COVID-19 infection.

**Abstract**

Radiological image is currently adopted as the visual evidence for COVID-19 diagnosis in clinical. Using deep models to realize automated infection measurement and COVID-19 diagnosis is important for faster examination based on radiological imaging. Unfortunately, collecting large training data systematically in the early stage is difficult. To address this problem, we explore the feasibility of learning deep models for COVID-19 diagnosis from a single radiological image by resorting to synthesizing diverse radiological images. Specifically, we propose a novel conditional generative model, called CoSinGAN, which can be learned from a single radiological image with a given condition, i.e., the annotations of the lung and COVID-19 infection. Our CoSinGAN is able to capture the conditional distribution of visual finds of COVID-19 infection, and further synthesize diverse and high-resolution (512×512) radiological images that match the input conditions precisely. Both deep classification and segmentation networks trained on synthesized samples from CoSinGAN achieve notable detection accuracy of COVID-19 infection. Such results are significantly better than the counterparts trained on the same extremely small number of real samples (1 or 2 real samples) by using strong data augmentation, and approximate to the counterparts trained on large dataset (2846 real images). It confirms our method can significantly reduce the performance gap between deep models trained on extremely small dataset and on large dataset, and thus has the potential to realize learning COVID-19 diagnosis from few radiological images in the early stage of COVID-19 pandemic. Our codes are made publicly available at https://github.com/PengyiZhang/CoSinGAN.

Key words: COVID-19 diagnosis, generative model, single radiological image, conditional distribution

# 1. Introduction

The highly contagious Coronavirus Disease 2019 (COVID-19), caused by the severe acute respiratory syndrome coronavirus 2 (SARS-CoV-2) virus [1][2][3], has spread rapidly across the world and millions of people has been infected. This surge in infected patients has overwhelmed healthcare systems in a short time. Due to the close contact with patients, many medical professionals have also been infected, further worsening healthcare situation. To date (May 19th 2020), COVID-19 has resulted in over 4.8 million infections and 310,000 deaths. Early detection of COVID-19 is significantly important to prevent the spread of this epidemic.

Reverse transcription polymerase chain reaction (RT-PCR) is the de facto golden standard for COVID-19 diagnosis [4][5]. However, the global shortage in RT-PCR test kits has severely limited the extensive detection of COVID-19. Meanwhile, the current clinical experience implies RT-PCR has a low sensitivity [6][7][8] especially in the early outbreak of COVID-19. That means multiple testing may be required to rule out the false negative cases [9], which may delay the confirmation of suspected patients and take up huge healthcare resources.

Since most patients infected by COVID-19 are initially diagnosed with pneumonia [10], radiological examinations, including computed tomography (CT) and X-ray, are able to provide visual evidence of COVID-19 infection for diagnosis and patient triage. Existing chest CT findings in COVID-19 infection [11] have implied that chest CT screening on patient at the early stage of COVID-19 presents superior sensitivity over RT-PCR [7] and even confirms the false negative cases given by RT-PCR [4]. Therefore, radiological examinations are currently used as parallel testing in COVID-19 diagnosis. However, as the number of infected patients dramatically increases, clinicians need to analyze radiographs repeatedly, which brings huge pressure to them. Therefore, there is an immediate need for developing methods for automated infection measurement and COVID-19 diagnosis based on radiological images to reduce the efforts of clinicians and accelerate the diagnosis process.

Many approaches, mostly using deep models, have been proposed for automated COVID-19 diagnosis based on chest CT[9] [12][13][14] or chest X-ray [10] [15], and have claimed notable detection accuracy of COVID-19 infection. However, the research of these approaches tends to lag slightly behind the outbreak of COVID-19 pandemic. It is probably because accumulating sufficient radiological images that are required to train deep models is difficult in the early stage of COVID-19 pandemic. To solve the dilemma of training deep models on insufficient training samples and realize automated COVID-19 diagnosis in the early stage, some researches resort to shallow network [10], prior knowledge [10], transfer learning [15][16], and data augmentation method based on generative adversarial network (GAN) [16][17]. However, these methods still require relatively large training dataset, and thus cannot respond immediately to the outbreak of COVID-19 pandemic. An ideal solution is to learn COVID-19 diagnosis from a single radiological image.

In this paper, we explore the feasibility of learning deep models for COVID-19 diagnosis from a single radiological image by resorting to synthesizing diverse radiological images. Specifically, we propose a novel conditional generative model, called CoSinGAN, which can be learned from a single radiological image with a condition, i.e., the annotations of the lung and COVID-19 infection. Inspired by SinGAN [18], we build CoSinGAN with a pyramid of GANs, each of which has a two-stage UNet-style [19][20] generator and is responsible for translating condition mask into radiological image at a different scale. Unlike SinGAN estimating the 'unconditional' distribution of a single nature image, our CoSinGAN is designed to capture the 'conditional' distribution of a single radiological image. Estimating conditional distribution from a single image is much more difficult. Because one should prevent the generators from being 'overfitted' to the single input condition, and meanwhile, need to 'overfit' them to the single training image as much as possible. Therefore, we design the two-stage generator at each scale of CoSinGAN to cooperate with the multi-scale architecture by progressively adding image details and enhancing the condition constraints. Besides, we introduce a mixed reconstruction loss and a hierarchical data augmentation module to train CoSinGAN smoothly. The mixed reconstruction loss consists of weighted pixel-level loss (WPPL), multi-scale feature-level VGG [21]loss, multiscale feature-level UNet [19] loss and multi-scale structural similarity (MS-SSIM) [22][23] loss. The mixed reconstruction loss is able to provide rich and stable gradient information for the optimization of generators. The hierarchical data augmentation module can produce data augmentation with different intensities for the training of two-stage generators at different scales, which facilitates the balance between fitting conditions and fitting images. Moreover, to enable CoSinGAN to generate diverse radiological images, we provide two effective approaches, including randomizing input conditions and fusing images of different modalities. Extensive ablation experiments strongly confirm the efficacy of our proposed methods. Compared to the popular pix2pix [20] model, our CoSinGAN is able to synthesize diverse and high-resolution (512×512) radiological images that match the input conditions and visual finds of the lung and COVID-19 infection more precisely. Both deep classification and segmentation networks trained on synthesized samples from CoSinGAN achieve notable detection accuracy of COVID-19 infection. Such results are significantly better than the counterparts trained on the same extremely small number of real samples (1 or 2 real samples) by using strong data augmentation, and approximate to the counterparts trained on large dataset (2846 real images). It confirms our method can significantly reduce the performance gap between deep models trained on extremely small dataset and on large dataset, and thus has the potential to realize learning COVID-19 diagnosis from few radiological images in the early stage of COVID-19 pandemic.

## 2. CoSinGAN

CoSinGAN consists of three key components, including multi-scale architecture with a pyramid of two-stage GANs, a mixed reconstruction loss, and a hierarchical data augmentation module.

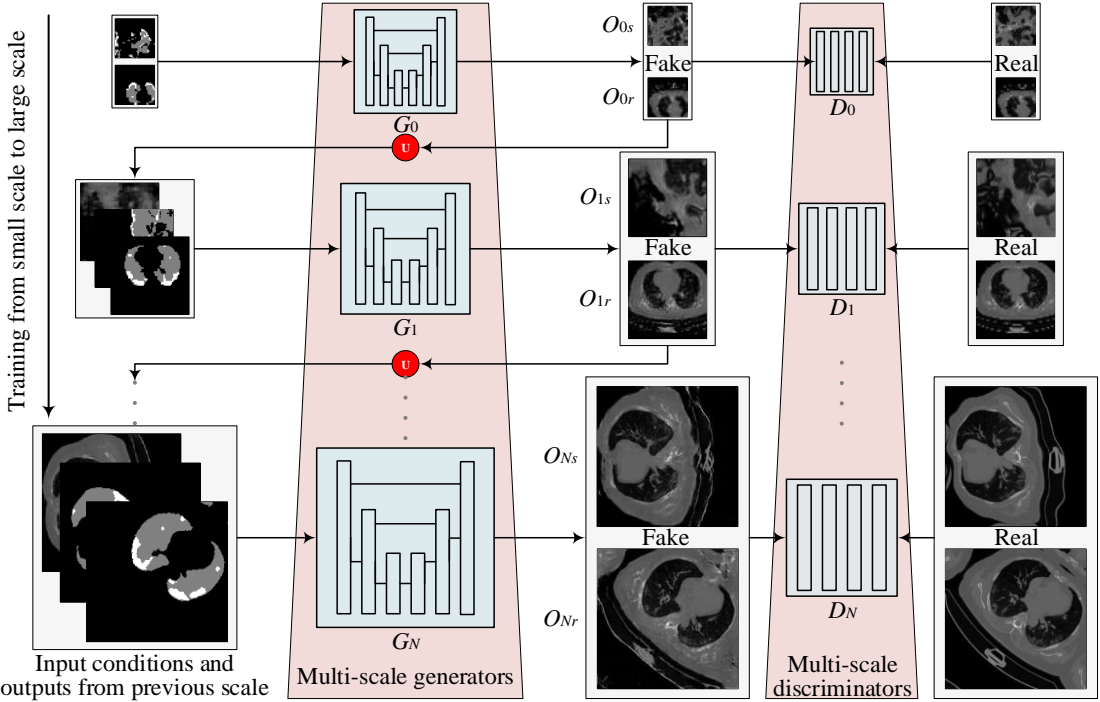

**Figure 2**. Multi-scale architecture of proposed CoSinGAN. CoSinGAN consists of a pyramid of GANs, each of which has a two-stage generator and is responsible for translating condition mask into radiological image at a different scale. The input to $G_i$ is an augmented condition mask, and the generated radiological image from the scale $i-1$, which is upsampled to the scale $i$ (except for scale 0). Through iterating optimizations from small image scale to large image scale, CoSinGAN progressively learns to generate realistic and high-resolution (512×512) radiological images with COVID-19 infection.

### 2.1 Multi-scale architecture with a pyramid of two-stage GANs

**Overall**. Learning a generative model to synthesize high-resolution and high-quality images is quite difficult due to the unstable training process of GAN. A useful trick is to learn a pyramid of GANs as adopted by SinGAN [18] to increase the resolution of generated images progressively. We borrow this trick and build CoSinGAN with a multi-scale architecture as depicted in **Fig 2**. It is worth noting that we expect to use the synthesized radiological images with COVID-19 infection to train both classification and segmentation models for COVID-19 diagnosis. Thus, the synthesized images should match the given input conditions precisely, especially in the infection regions. To achieve that, CoSinGAN is designed to capture the 'conditional' distribution of a single radiological image rather than the 'unconditional' distribution of a single nature image as done by SinGAN. Estimating conditional distribution from a single image is much more difficult, because one should pay more attention to preventing deep models from being 'overfitted' to the single input condition, and meanwhile, need to 'overfit' them to the single training image as much as possible. To tackle this problem, at each scale we design a two-stage GAN to cooperate with the pyramid hierarchy as illustrated in **Fig. 3**. The first stage is mainly responsible for fitting the input condition and increasing the resolution of radiological image, and the second stage is responsible for restoring image details that may not be reconstructed in the first stage. Through iterative optimization of enhancing condition constraints and restoring image details across all scales of GANs, our CoSinGAN is able to generate realistic and high-resolution radiological images that match the given input conditions precisely as demonstrated in **Fig. 1** and **Fig. 9**.

**Multi-scale architecture**. As shown in **Fig. 2**, CoSinGAN consists of $N+1$ GANs, i.e., multi-scale generators $\{G_0, G_1, ..., G_N\}$ and multi-scale discriminators $\{D_0, D_1, ..., D_N\}$, corresponding to $N+1$ different image scales. The original single training image $X_{orig}$ and its condition mask $C_{orig}$ are initially resized to the pre-defined image scales respectively to conduct the training sample $\{X_i, C_i | i \in [0, N]\}$ for $N+1$ GANs. The training of CoSinGAN starts from the coarsest image scale 0, and gradually passes through all image scales. For the generation of radiological images at the specific scale $i$, the generators $\{G_0, G_1, ..., G_i\}$ are sequentially involved, where the output $O_{j-1}$ of $G_{j-1}$ is upsampled to the $j$-th image scale, and is further combined with $C_j$ to build the input of $G_j$ ($j \in [1, i]$). Benefiting from the output $O_{j-1}$ of previous generator $G_{j-1}$, $G_j$ will not fail quickly in the adversarial learning of GAN, and will continue to fight with $D_j$ and learn to generate realistic radiological image gradually. As there is no previous scale, $G_0$ learns to map the conditional mask $C_0$ into radiological image directly. Due to the small image scale, $G_0$ can be trained easily and further start the training of subsequent image scales smoothly. The multi-scale conditions $\{C_i | i \in [1, N]\}$, modulating the input of GANs across $N$ image scales, will enforce the output of CoSinGAN to match the given conditions strictly, which is exactly what we expect.

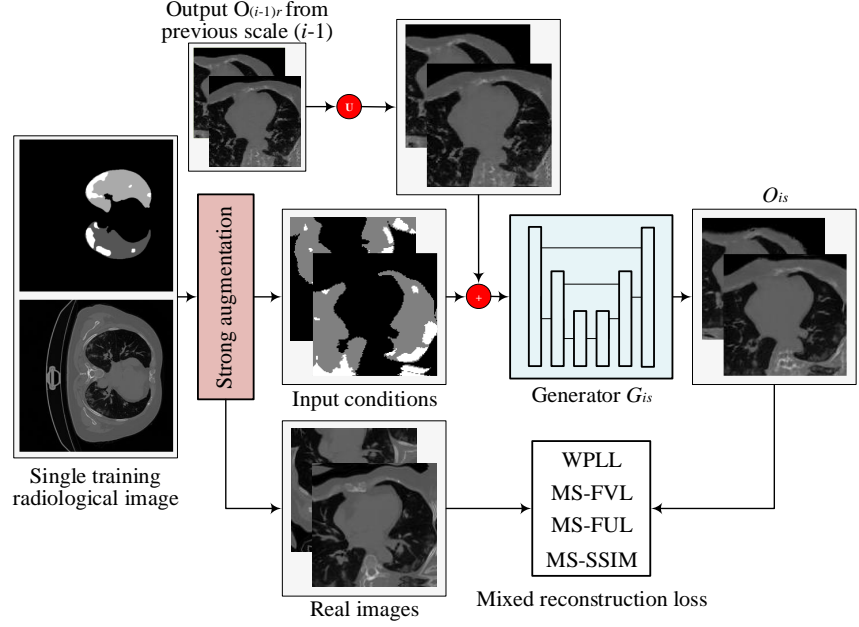

(a) The generator $G_{is}$ with a conditional image super-resolution framework

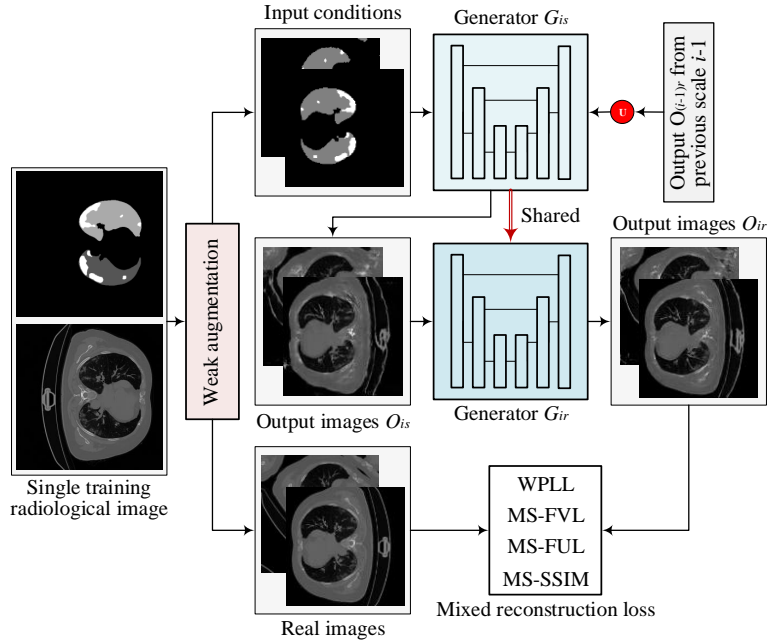

(b) The generator $G_{ir}$ with an unconditional image restoration framework

**Figure 3**. Illustration of the architecture of proposed two-stage GAN at the $i$-th image scale.

**Two-stage GAN**. At a specific image scale $i$, we design a two-stage GAN as depicted in **Fig. 3**. The generator $G_i$ in its first stage, called $G_{is}$, is designed to perform conditional image super-resolution, responsible for learning the condition constraints and increasing the resolution of radiological image simultaneously. $G_{is}$ inputs the output $O_{i-1}$ from previous scale $i$-1 that is first upsampled to the $i$-th image scale and sequentially modulated by the condition $C_i$, and outputs an image $O_{is}$ with the $i$-th image scale

$$O_{is} = G_{is}\left(U_i\left(O_{i-1}\right) + C_i\right), \qquad (1)$$

where $U_i$ denotes the upsampling operation. In the second stage, we directly copy the trained generator $G_{is}$ to perform unconditional image restoration, thus called $G_{ir}$. $G_{ir}$ inputs $O_{is}$ and outputs an image $O_{ir}$ with the $i$-th image scale

$$O_{ir} = G_{ir}\left(O_{is}\right), \qquad (2)$$

Thus, the full image generation process of proposed two-stage GAN can be formulated as

$$O_i = G_{ir}\left(G_{is}\left(U_i(O_{i-1}) + C_i\right)\right), \quad (3)$$

We specially design a hierarchical data augmentation module, which can produce strong augmentation and weak augmentation (detailed in section 2.3) to train such a two-step GAN. We first perform strong augmentation on training sample $(X_i, C_i)$ to train $G_{is}$ and thus ensure $G_{is}$ can generalize to different input conditions despite of the possibility of blurring the generated image. Whereas we do weak augmentation on training sample $(X_i, C_i)$ to train $G_{ir}$ and thus facilitate $G_{ir}$ to restore image details as much as possible despite of the possibility of violating the given conditions. Therefore, the two-stage GAN is actually trained by a two-step optimization: (a) enhance given condition constraints but may blur image details, and (b) restore image details but may violate given conditions. By iterating such a two-step optimization through all image scales progressively, the two-stage GANs with larger image scales are able to generate high-resolution radiological images that match the given conditions strictly, and meanwhile, have clear and accurate image details.

**Implementation details**. A total of 9 image scales are used in our implementation of CoSinGAN for synthesizing high-resolution chest CT slices, including 32×32, 48×48, 64×64, 96×96, 128×128, 192×192, 256×256, 384×384 and 512×512. We purposely choose such image scales to facilitate the design of the multi-scale generators with different numbers of down-sampling layers. Specifically, we choose a network architecture similar with the popular pix2pix model [20], including a UNet-style generator and a patch discriminator. Considering the reusability of trained models between two adjacent image scales, we set the number of 2×downsampling layers in the UNet-style generators of CoSinGAN to (4, 4, 5, 5, 6, 6, 7, 7, 8), respectively. Meanwhile, the numbers of convolutional layers in discriminators are set accordingly to (6, 6, 7, 7, 8, 8, 9, 9, 10).

**2.2 Objective**

At the $i$-th image scale, we train generator $G_i$ in the manner of adversarial learning to obtain realistic images. It is done by learning $G_i$ to minimize the reconstruction loss $\ell_{rec}$ and the adversarial loss $\ell_{adv}$ simultaneously, thereby fooling the discriminator $D_i$ to maximize the probability of generated image being classified as real image. Therefore, our objective for optimizing $G_i$ is

$$\min_{G_i} \ell_{adv}\left(D_i\left(C_i, G_i(C_i, O_{i-1})\right), 1\right) + \ell_{rec}\left(C_i, G_i(C_i, O_{i-1}), X_i\right), \quad (4)$$

and the objective for optimizing $D_i$ is

$$\min_{D_i} \frac{1}{2}\left(\ell_{adv}\left(D_i\left(C_i, G_i(C_i, O_{i-1})\right), 0\right) + \ell_{adv}\left(D_i(C_i, X_i), 1\right)\right), \quad (5)$$

The same adversarial loss $\ell_{adv}$ as pix2pix [20] is adopted in our implementation. Besides, we propose a mixed reconstruction loss, including the weighted pixel-level loss (WPPL) $\ell_{WPPL}$, multi-scale feature-level VGG [21] loss (MS-FVL) $\ell_{MS-FVL}$, multi-scale feature-level UNet [19] loss (MS-FUL) $\ell_{MS-FUL}$ and multi-scale structural similarity (MS-SSIM) loss [22][23] $\ell_{MS-SSIM}$:

$$\ell_{rec} = \lambda_{WPPL} \times \ell_{WPPL} + \lambda_{MS-SSIM} \times \ell_{MS-SSIM} + \lambda_{MS-FVL} \times \ell_{MS-FVL} + \lambda_{MS-FUL} \times \ell_{MS-FUL}, \quad (6)$$

where $\lambda_{WPPL}$, $\lambda_{MS-SSIM}$, $\lambda_{MS-FVL}$ and $\lambda_{MS-FUL}$ denote the loss weights of WPPL, MS-SSIM, MS-FVL, and MS-FUL, respectively. Such a mixed reconstruction loss is able to provide rich and stable gradient information for the optimization of generators.

**WPPL**. WPPL computes the weighted mean of L1 distances between the pixels of generated image and real image, where the weight of each pixel is determined by its category, i.e., background, lung or COVID-19 infection,

$$\ell_{WPPL}(C_i, O_i, X_i) = \frac{1}{P}\sum_{p=1}^{P} M(C_i^p)|O_i^p - X_i^p|, \quad (7)$$

where $p$ is the pixel index, $P$ is the total number of pixels and M denotes a mapping function from category to weight. We use L1 loss rather than mean squared error (MSE) loss because optimizing MSE loss tends to obtain over-smoothed image details. We suggest a relatively higher weight for the regions of the lung and COVID-19 infection to emphasize the reconstruction of the lung and COVID-19 infection.

**MS-SSIM loss**. Different from mean-based metrics like L1 distance and MSE, SSIM [24] and MS-SSIM [22] are perceptually motivated metrics to evaluate image similarity based on local structure. As discussed in [23], MS-SSIM loss is differentiated and thus can be back-propagated to optimize the parameters of CoSinGAN. We adopt MS-SSIM loss [23] to optimize the reconstruction of local anatomical structures.

**MS-FVL**. The distance between deep features of two images from a pre-trained CNN classifier is frequently used as the perceptual loss [25][26][27] in image restoration tasks. Compared with pixel-level metrics, perceptual loss is able to obtain visually appealing results. The multi-scale feature-level VGG loss [27] used at the $i$-th scale of CoSinGAN is formulated as:

$$\ell_{MS-FVL}(O_i, X_i) = \sum_{j=1}^{J} \eta_j \frac{1}{P_j} \left\| F_j(X_i) - F_j(G_i(C_i, O_{i-1})) \right\|^1, \quad (8)$$

where $F_j$ denotes the $j$-th layer with $P_j$ elements of the VGG network [21] and $\eta_j$ denotes the weight of the $j$-th feature scale.

**MS-FUL**. Similar with MS-FVL, we design a multi-scale feature-level UNet loss, which measures the similarity of two images using the deep features from a pre-trained UNet [19]:

$$\ell_{\text{MS-FUL}}(O_i, X_i) = \sum_{k=1}^{K} \gamma_k \frac{1}{P_k} \left\| F_k(X_i) - F_k(G_i(C_i, O_{i-1})) \right\|^1, \quad (9)$$

where $F_k$ denotes the $k$-th layer with $P_k$ elements of the UNet network [19] and $\gamma_k$ denotes the weight of the $k$-th feature scale. Compared to VGG features that are trained for classification tasks, the UNet features trained for semantic segmentation encode much more positional and structural information, and thus are more sensitive to the positional distribution of pixels.

**2.3 Hierarchical data augmentation**

As described in section 2.1, to learn the conditional distribution from one single image, one need to well handle the two things: (a) ensure the generator can generalize to different input conditions, and (b) fit the single image as much as possible for visually accurate and appealing results. Performing strong data augmentation on the single training image is an effective approach to avoid overfitting, whereas it may corrupt the real data distribution and put an additional burden on the generator, and thus lead to blurry and unreal images. It is critical to design an appropriate data augmentation module to tackle this problem. Accordingly, we propose a hierarchical data augmentation module, involving strong augmentation and weak augmentation, to collaborate with the proposed two-stage GANs at multiple image scales. Specifically, at the $i$-th image scale, the hierarchical data augmentation module produces strong augmentation (SA) to train $G_{is}$, and produces weak augmentation (WA) to train $G_{ir}$. Meanwhile, as the image scale increases, the intensity of SA decreases gradually whereas WA keeps unchanged. Some augmented images and conditions produced by the hierarchical data augmentation module are visualized in **Fig. 4**.

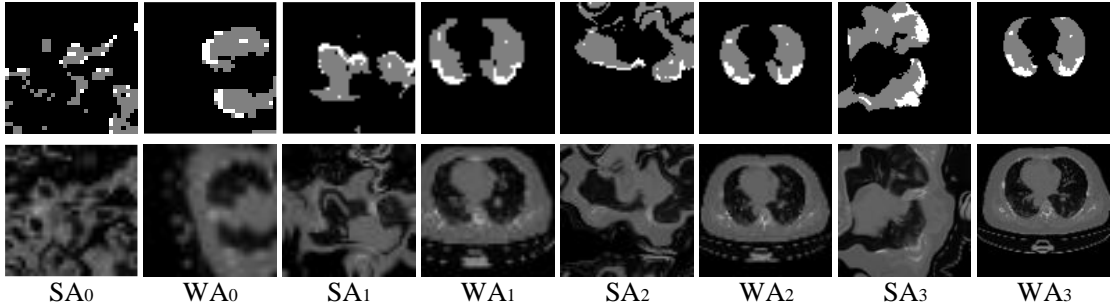

SA$_0$    WA$_0$    SA$_1$    WA$_1$    SA$_2$    WA$_2$    SA$_3$    WA$_3$

**Figure 4**. Illustration of augmented images and conditions produced by the hierarchical data augmentation module. SA and WA denote strong augmentation and weak augmentation, respectively. The subscript represents the index of image scale (images are resized for better visualization).

We design such a hierarchical data augmentation module with following advantages:
(1) SA is very critical to enable $G_{is}$ to generalize to different input conditions.
(2) WA can facilitate to fit the real image distribution without introducing additional learning burden.
(3) Decreasing the intensity of SA along with the increasing of image scales can well handle the balance between fitting conditions and fitting images.

Specifically, we implement the hierarchical data augmentation module based on random cropping, random rotation, random horizontal flipping, random vertical flipping and elastic transform. SA is designed by composing all these transforms, where the cropping size is between 0.5 and 1 times the image size and the parameters of elastic transform is set according to the specific image size. In comparison, WA does not use elastic transform, and the cropping size is between 0.75 and 1 times the image size. It is worth noting that the augmentation imposed on images should be consistent with augmentation imposed on conditions all the time. At the $i$-th image scale, to obtain consistent input for the generate $G_i$, we perform SA or WA on conditions and use the augmented conditions to generate the previous output $O_{i-1}$ from scale 0 to scale $i$-1 rather than directly imposing SA or WA on $O_{i-1}$ that is generated by original conditions. Besides, we treat the augmented samples as different samples and thus construct batched samples to realize mini-batch training.

## 3. Experiments and results

In this paper, we explore the feasibility of learning COVID-19 diagnosis from a single radiological image. We resort to synthesizing diverse radiological images with COVID-19 infection and thus propose a novel conditional GAN, i.e., CoSinGAN, to realize the radiological image generation process. Therefore, we first conduct experiments to evaluate the effectiveness of CoSinGAN on synthesizing high-resolution and high-quality radiological images that can well match the given conditions and visual finds of the lung and COVID-19 infection. Next, we evaluate the effectiveness of synthesized radiological images from CoSinGAN on training both classification and segmentation networks for COVID-19 diagnosis.

## 3.1 Materials

We use the public COVID-19-CT-Seg dataset [28], which consists of 20 public COVID-19 CT scans with pixel-level annotations of the left lung, right lung and COVID-19 infection. The annotations, first labeled by junior annotators, are refined by two radiologists with 5 years experience, and are further verified and refined by a senior radiologist with more than 10 years experience in chest radiology. In our experiment, we randomly select 15 scans for training and the other 5 scans are left for test. We slice them into slices and resize these slices to the size of 512×512, and thus constitute a training set with 2846 chest CT slices and a test set with 674 chest CT slices, respectively. We observe that the CT slices in our materials mainly present two distinct appearances as shown in **Fig. 5,** which may be caused by using different ranges of Hounsfield Unit during the CT imaging process. For convenience, we roughly treat them as two different modalities, and name them modality 1 and modality 2 respectively to indicate the difference.

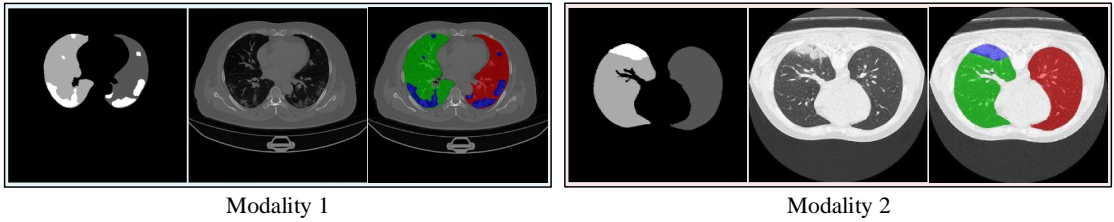

Modality 1    Modality 2

**Figure 5**. Illustration of the material used in our experiments.

## 3.2 Experiments and results on synthesizing radiological images

We use two representative slices of different modalities as depicted in **Fig. 5** from the training set to train two individual CoSinGANs separately. We first conduct ablation experiments on the three key components of CoSinGAN to verify their efficacies. Second, we perform evaluation and comparison on the image quality of synthesized radiological images. We finally conduct experiments to test the ability of CoSinGAN in generating diverse samples.

### 3.2.1 Training details

We train CoSinGAN with 9 image scales, including 32×32, 48×48, 64×64, 96×96, 128×128, 192×192, 256×256, 384×384 and 512×512, which means 9 two-stage GANs are required to train sequentially from coarsest scale to finest scale. The loss weights of WPPL, MS-SSIM, MS-FVL and MS-FUL in proposed mixed reconstruction loss are empirically set to 10.0, 1.0, 10.0 and 10.0. As suggested in section 2.2, we set the category weights of background, lung and COVID-19 infection in WPPL to 0.1, 0.5 and 1.0 respectively to emphasize the reconstruction of the lung and COVID-19 infection. Meanwhile, we set the pixel values of these three categories in the input conditions to 0, 128 and 255, separately. We do strong augmentation to train these two-stage GANs with 4000 epochs and mini-batch of 4 in their first stage by using Adam optimizer with the parameters of $\beta_1 = 0.5$ and $\beta_2 = 0.999$. We use an initial learning rate of 0.0002 that is linearly decayed by 0.05% each epoch after 2000 epochs. Correspondingly, we perform weak augmentation to train these two-stage GANs with 2000 epochs and mini-batch of 4 in their second stage by using Adam optimizer with the parameters of $\beta_1 = 0.5$ and $\beta_2 = 0.999$. We use an initial learning rate of 0.0001 that is linearly decayed by 0.1% each epoch after 1000 epochs. The batch size of training samples at the image scale of 512×512 is set to 2 due to memory limitation. The trained models are further used to synthesize radiological images with given input conditions from the training set of our materials.

### 3.2.2 Ablation experiments

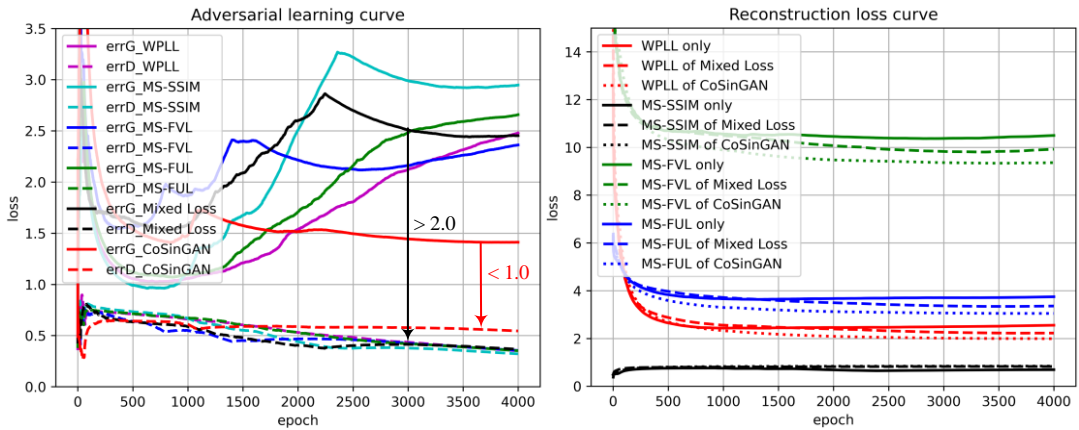

**Figure 6**. The learning curves of CoSinGAN trained with different reconstruction loss functions at the image scale of 256×256. We plot the moving average of loss values for better visualization.

**Mixed reconstruction loss**. We introduce mixed reconstruction loss to provide rich and stable gradient information for the optimization of generators. To evaluate its efficacy, we train CoSinGAN with a single scale of $256 \times 256$ on the single training image of modality 1 by adopting WPLL, MS-SSIM loss, MS-FVL, MS-FVL, and mixed reconstruction loss as the reconstruction loss function separately. The training curves, including adversarial learning curves and reconstruction loss curves, are depicted in **Fig. 6**. We directly use the first stage of the two-stage GAN to perform our evaluation, thus simplifying CoSinGAN to have a similar architecture with the well-known pix2pix model [20]. These trained models are used to generate radiological images with given input conditions as shown in **Fig. 7**.

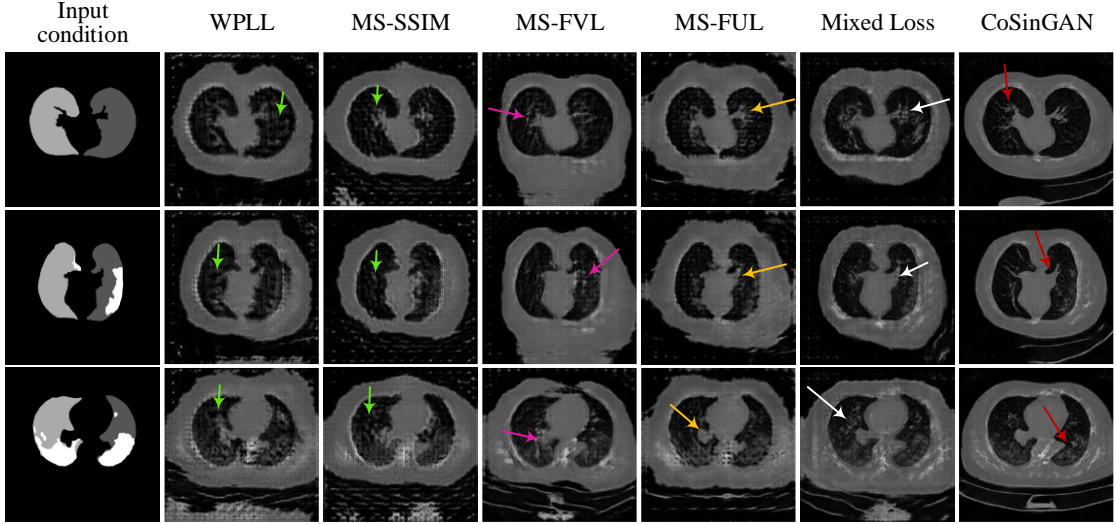

**Figure 7**. Comparison of the synthesized images with image size of $256 \times 256$ from CoSinGAN trained with different reconstruction loss. Arrows are used to highlight the differences.

Compared to WPLL and MS-SSIM loss, MS-FVL using deep features from pre-trained VGG network tends to produce visually pleasing images with less noise, but at the cost of losing more local image details as indicated by pink arrows in the fourth column of **Fig. 7**. MS-FUL also achieves visual impact similar with MS-FVL, but can reconstruct more image details like sharp contours and edges (highlighted by yellow arrows in the fifth column of **Fig. 7**) than MS-FVL. We argue it is because that the deep features from UNet, designed for semantic segmentation, encode much more positional and structural information, and thus make MS-FUL sensitive to the positional distribution of pixels. Correspondingly, WPLL and MS-SSIM loss, using raw pixel features, can synthesize much more image details as pointed by green arrows in **Fig.7** at the cost of presenting visual unpleasing impact with more background noises. By combining WPLL, MS-SSIM loss, MS-FVL and MS-FUL together, our mixed reconstruction loss can inherit advantages of them, complement with each other, and thus produce visual pleasing images with less noise and more local details (highlighted by white arrows). Moreover, as depicted in **Fig. 6** (b), the model trained with mixed reconstruction loss achieves consistently smaller WPLL, MS-FVL and MS-FUL than the same model trained with only one of loss items in the mixed reconstruction loss. It strongly confirms the mutual collaboration efficacy between different loss items in the mixed reconstruction loss.

**Multi-scale architecture and two-stage GAN**. We train a complete CoSinGAN with 9 two-stage GANs on the single training image of modality 1. We plot the adversarial learning curve and reconstruction loss curve of CoSinGAN with the image scale of 256×256 in **Fig. 6**. It is worth noting that all the models compared in **Fig. 6** use the same training configuration except that the complete CoSinGAN is trained gradually from the scale of 32×32 to the scale of 256×256. As can be seen from **Fig. 6** (a), the complete CoSinGAN presents a better adversarial learning curve than the other models trained with a single scale. The adversarial loss values of generator $G$ and discriminator $D$ are kept close to each other throughout the entire training process. It indicates that the adversarial training of CoSinGAN is stable and thus $G$ is able to capture the distribution of real images gradually through the continuously fighting with $D$. The reconstruction loss curve in **Fig. 6** (b) also shows that the complete CoSinGAN trained with multi-scale architectures achieves lower fitting error. Besides, the radiological images produced by the complete CoSinGAN as depicted in **Fig. 7** present significantly better visual impact with realistic and sharp image details (highlighted by red arrows). These results strongly verify the effectiveness of multi-scale architecture. Moreover, we use the complete CoSinGAN to generate images with all 9 scales and compare them in **Fig. 8**. Each scale includes two synthesized images, one from the first stage and the other from the second stage. We first use the red arrows to track the contour of lung and thus highlight the efficacy of proposed multi-scale architectures and two-stage GAN on enhancing the condition constraints. We can easily identify that the contour of lung in synthesized radiological images gradually match the input conditions. Besides, we utilize the green arrows to track the details of lung and COVID-19 infection in synthesized images as the image scale increases. Intuitively, the image details are enhanced progressively. Such results strongly confirm our claims that our multi-scale architectures are able to collaborate with the two-stage GANs by iteratively enhancing conditions and details.

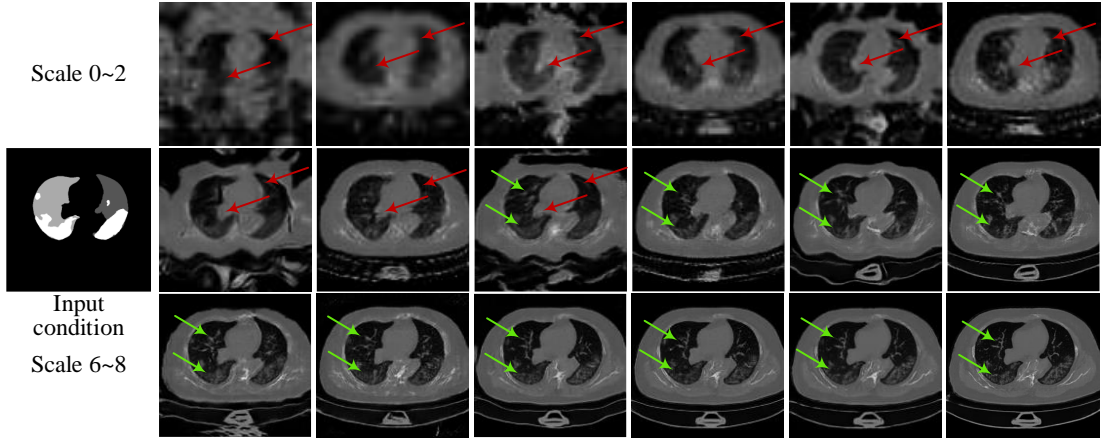

**Figure 8**. Illustration of the synthesized images with all 9 scales from CoSinGAN. Each scale consists of two synthesized images, one from the first stage and the other from the second stage. All images are resized to $512\times512$ for better visualization.

**Hierarchical data augmentation**. As depicted in **Fig. 4**, our hierarchical data augmentation module is able to produce strong augmentation (SA) for the first stage of GAN and weak augmentation (WA) for the second stage of GAN. SA is designed to enhance conditions, while WA is used to facilitate the restoration of image details. As can be seen from the first three scales of synthesized images in **Fig. 8**, the generators trained with SA in the first stage present strong generalization ability to input condition because the shape of lung is synthesized consistently with the input condition. The generators trained with WA in the second stage show weak generalization ability to input condition as the shape of lung is not well maintained. Besides, the synthesized images from the second stage tend to be more realistic and have more details than those from the first stages. As the image scale increases, the intensity of SA gradually decreases, whereas the output of CoSinGAN contains more and more condition information. Benefiting from the output from previous scales, the generators in the later scales trained with relatively weaker SA are still able to generalize to input conditions. Thus, more energy can be assigned to optimize image details. Such results clearly confirm that our hierarchical data augmentation module is able to provide a well balance between preventing the generator from being overfitted to input condition and facilitating to overfit the single training image as much as possible when learning conditional distribution from a single image.

### 3.2.3 Evaluation and comparison on image quality

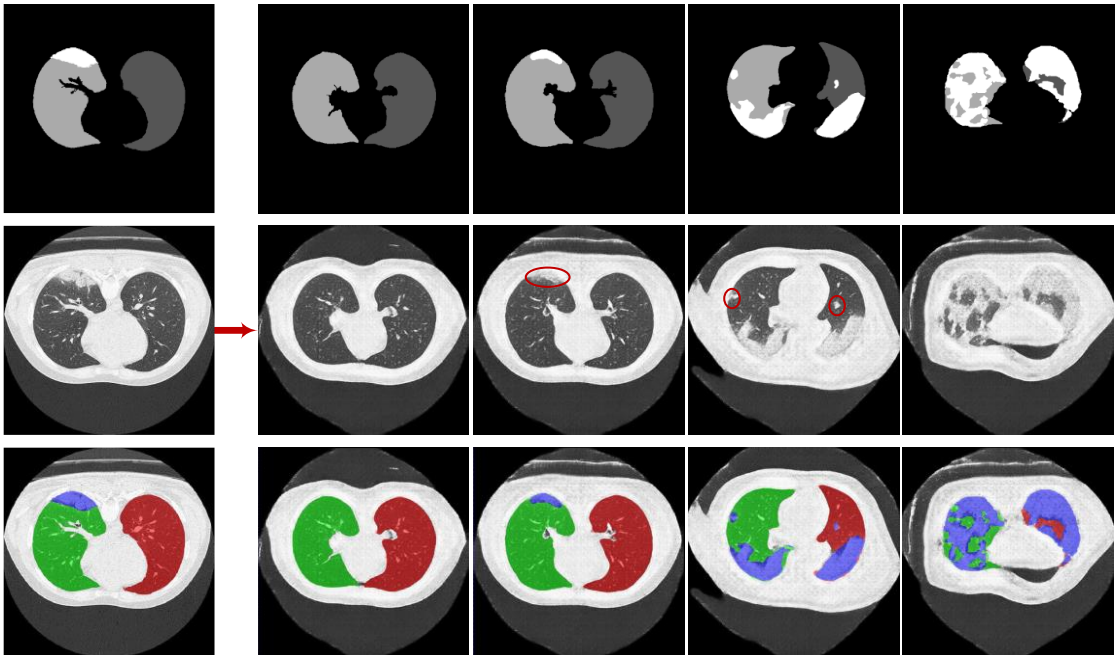

**Figure 9**. The radiological images of modality 2 synthesized by proposed CoSinGAN with given conditions.

**Baselines**. The pix2pix model [20] is a well-known conditional GAN framework for image-to-image translation. In our implementation, we directly adopt the enhanced pix2pix model to build our two-stage GAN of CoSinGAN. Specifically, we replace L1 reconstruction loss in the pix2pix model with the proposed mixed reconstruction loss to obtain the enhanced pix2pix model. Accordingly, we use the pix2pix model and the enhanced pix2pix model as our baselines to compare with our CoSinGAN and highlight our contributions. It is worth noting the baseline pix2pix model is also implemented with weighted L1 reconstruction loss, i.e., WPLL, to emphasize the reconstruction of lung and COVID-19 infection. We first train these models on the single image of modality 1 by using the training setting detailed in **section 3.2.1** and sequentially train these models on the single image of modality 2.

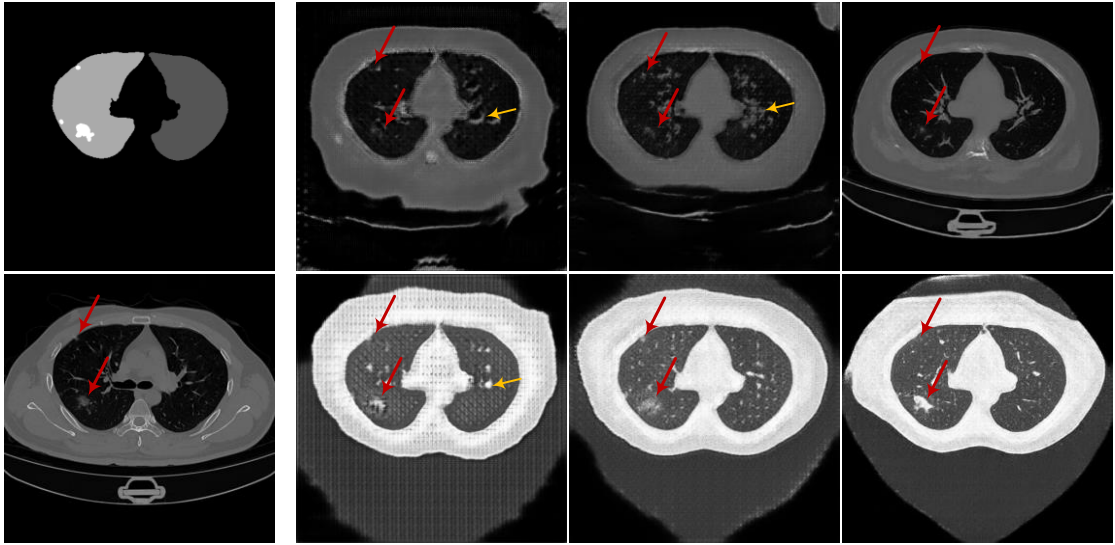

**Figure 10**. The synthesized radiological images of modality 1 and modality 2 with given the same input condition. The input condition and the reference ground-truth radiological image are depicted in the first column. The last three columns are the results of pix2pix, enhanced pix2pix and CoSinGAN respectively, where the top is modality 1 and the bottom is modality 2 in each column. Red arrows are used to track and highlight the small regions of COVID-19 infection.

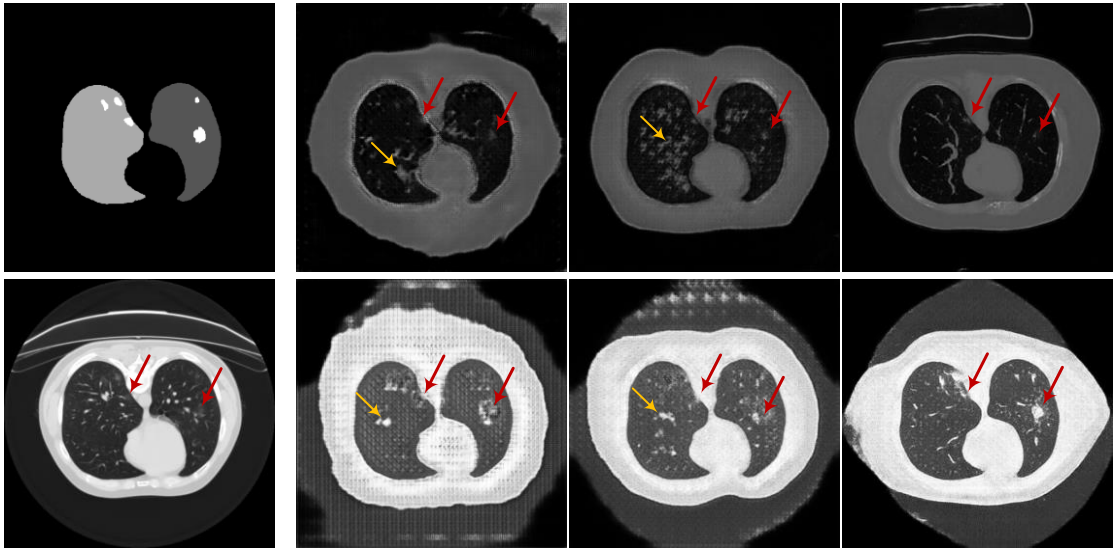

**Figure 11**. The synthesized radiological images of modality 1 and modality 2 with given the same input condition. The input condition and the reference ground-truth radiological image are depicted in the first column. The last three columns are the results of pix2pix, enhanced pix2pix and CoSinGAN respectively, where the top is modality 1 and the bottom is modality 2 in each column. Red arrows are used to track and highlight the small regions of COVID-19 infection.

**Qualitative comparison**. We first show the synthesized images of modality 1 and modality 2 from our CoSinGANs in **Fig. 1** and **Fig. 9** to give a direct visual impact. As can be seen, CoSinGAN is very sensitive to the input conditions as even the small isolated regions of COVID-19 infection can be reconstructed very well (highlighted by red circles). Meanwhile, these synthesized radiological images are able to present sharp and rich image details with low noise and clean background, comparable to the single training image with the size of 512×512. The visual appearance of the lung and COVID-19 infection is also synthesized consistently with the training image. Such property of CoSinGAN is very important, and can facilitate the construction of synthesized training samples with pixel-level annotations of the lung and COVID-19 infection to explore the feasibility of learning COVID-19 diagnosis from a single radiological image. We then compare the results of different models with given the same input conditions in **Fig. 10**, **11** and **12**. These input conditions are sampled from different CT scans, where their corresponding real images may present different modalities, called reference ground-truth images. As can be seen, our CoSinGAN can produce visually appealing results with clear image details and clean background, significantly better than the baselines and comparable to the reference ground-truth images. First, the results of the pix2pix model contain too much grid artifacts, thus leading to visually unpleasant results. Meanwhile, the synthesized details of lungs are not clearly and appear to be lung artifacts, which makes it very difficult to distinguish the synthesized COVID-19 infection from these artifacts (highlighted by yellow arrows). Next, benefiting from the mixed reconstruction loss, the enhanced pix2pix model achieves better visual impact with less grid artifacts and richer lung details compared to pix2pix. Despite that, the synthesized details of lungs are also not clear enough to be distinguished from the synthesized COVID-19 infection (indicated by yellow arrows in the third column). Such synthesized images of pix2pix model and enhanced pix2pix model cannot be used to learn COVID-19 diagnosis smoothly. In comparison, our CoSinGAN effectively solves the problems of grid artifacts and blurred lung details, and can produce high-quality radiological images with accurate details of the lung and COVID-19 infection, facilitating to train deep models for COVID-19 diagnosis. We finally compare our results with the reference ground-truth images and our results achieve comparable image sharpness. We also find that the COVID-19 infection regions in the reference ground-truth images present different but correlated visual appearances in different CT scans, which motivates us to test the ability of CoSinGAN in synthesizing diverse radiological images with COVID-19 infection.

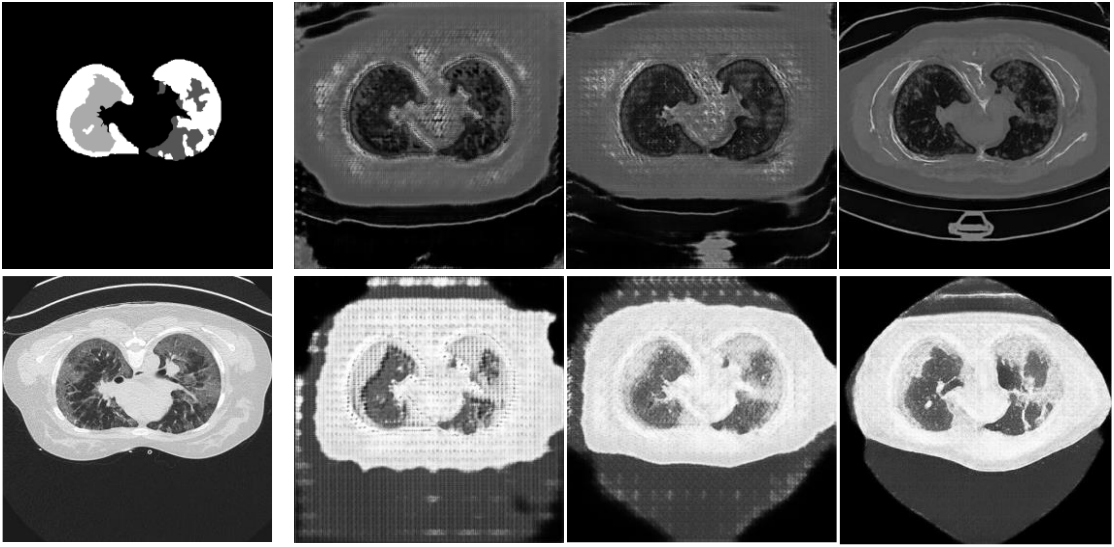

**Figure 12** The synthesized radiological images of modality 1 and modality 2 with given the same input condition. The input condition and the reference ground-truth radiological image are depicted in the first column. The last three columns are the results of pix2pix, enhanced pix2pix and CoSinGAN respectively, where the top is modality 1 and the bottom is modality 2 in each column.

**Table 1**. Image quality scores of synthesized radiological images. Our CoSinGAN surpasses the baseline methods by a large margin.

| Models | Modality 1 | | | | Modality 2 | | | | Overall | |
|---|---|---|---|---|---|---|---|---|---|---|
| | ENet | | UNet | | ENet | | UNet | | | |
| | Lung | Infection | Lung | Infection | Lung | Infection | Lung | Infection | Lung | Infection |
| Pix2pix | 90.7 | 74.5 | 91.1 | 59.2 | 71.5 | 32.3 | 63.9 | 51.1 | 79.3 | 54.3 |
| Enhanced Pix2pix | 73.4 | 75.7 | 87.8 | 55.7 | 68.9 | 49.7 | 68.2 | 51.0 | 74.6 | 58.0 |
| CoSinGAN | 73.8 | 79.6 | 92.3 | 87.8 | 85.3 | 50.1 | 90.7 | 73.7 | 85.5 | 72.8 |

**Quantitative comparison**. To quantify the quality of the generated radiological images, we follow the same evaluation method in [20][27]. We use the baseline segmentation networks, i.e., ENet [29] and UNet [19] (detailed later in section 3.3), that are well trained for the lung and COVID-19 infection segmentation on the training set of our materials to segment the synthesized images, and compare how well the segmentation outputs match the corresponding inputs of CoSinGAN. The intuitive is that if CoSinGAN can produce realistic radiological images, the segmentation networks trained on real images should be able to well segment them. The common Dice similarity coefficient (DSC) is computed as the image quality score to compare different models. Specifically, we first use pix2pix, enhanced pix2pix and CoSinGAN to synthesize the same number (2846) of radiological images with all given conditions in the training set of our materials. We then perform lung and COVID-19 infection segmentation on these synthesized images, and calculate the mean DSC scores of the lung and COVID-19 infection as the image quality scores to compare different models. As detailed in **Table 1**, our CoSinGAN obtains the highest image quality scores on both lung and COVID-19 infection, surpassing the baseline methods by a large margin. It indicates that CoSinGAN has reconstructed the visual appearance of the lung and COVID-19 infection more precisely at the locations specified in the input conditions. Such results strongly confirm the effectiveness of our CoSinGAN on learning the conditional distribution of radiological image from a single radiological image.

### 3.2.4 Evaluation on the ability of CoSinGAN in generating diverse samples

The images synthesized by GAN tends to be lack of diversity and present a single modality, which does not facilitate the training of deep models. Given an input condition, we expect CoSinGAN is able to generate diverse samples that are different but correlated in visual appearance. We explore three approaches to enable such ability of CoSinGAN, including applying dropout at test time, randomizing the input condition and fusing synthesized images of different modalities, called data diversification methods. We expect to use such methods to improve the diversity of synthesized radiological images and thus enable deep models to be trained effectively on synthesized radiological images for COVID-19 diagnosis.

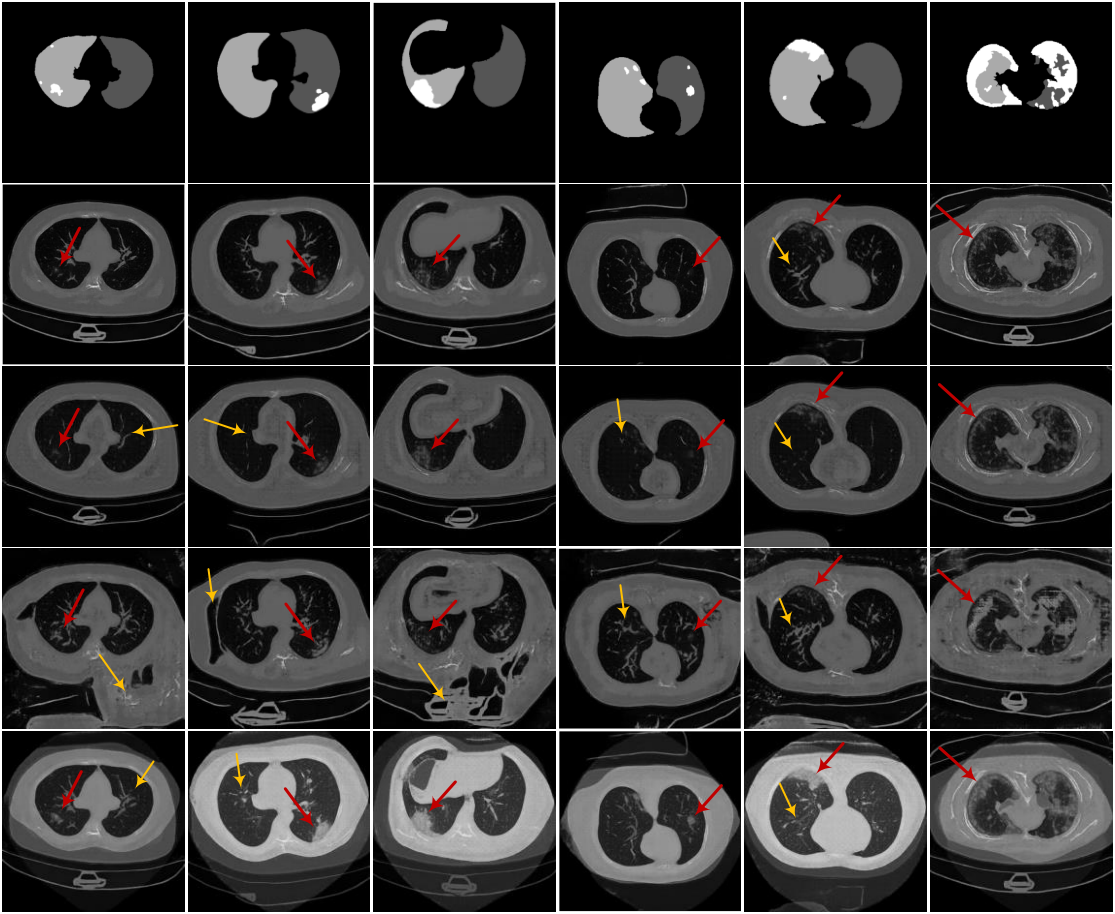

**Figure 13**. The diversification results of synthesized radiological images. The rows from top to bottom represent input conditions, synthesized images without using any data diversification methods, results of applying dropout at test time, results of randomizing input condition and results of fusing synthesized images of different modalities, respectively. Red and yellow arrows highlight the differences between diversification results.

**Applying dropout (AD)**. Applying dropout at inference time with a dropout rate of 50% can add randomness to the forward propagation of CoSinGAN through randomly inactivating some activation units of neural network. As can be seen, this operation presents a slight effect on CoSinGAN's output, including the fade of some image details (indicated by arrows in the third row of **Fig. 13**). Thus, this approach may not contribute much to the diversity of synthesized radiological images.

**Randomizing input condition (RC)**. During the training process, the pixel values of background, lung and COVID-19 infection in the input conditions are set to 0, 128 and 255, separately. After training, we can randomize the input condition by adding random noise to it to synthesize diverse images. Specifically, the pixel values of background, lung and COVID-19 infection are randomly set as follows:

$$V_{background} \in [0, \delta_b], V_{lung} \in [128-\delta_l, 128+\delta_l], V_{infection} \in [255-\delta_i, 255], \quad (10)$$

where $\delta_b$, $\delta_l$, and $\delta_i$ denote the magnitudes of the random noises. In our experiment, we set them to 16,

16 and 32, respectively. It is worth noting that such randomness also exists in the multi-scale input conditions. That means the input condition at each image scale may be different, which further promotes the synthesis of diverse samples. As can be seen, RC produces relatively diverse radiological images with notable differences in background, local lung details and COVID-19 infection (highlighted by the arrows in the fourth row of **Fig. 13**). Besides, RC does not spoil the sharpness and local details of synthesized images. Despite the lack of clinical evidences for the diverse appearance of synthesized COVID-19 infection, such results still confirm that RC is an effective data diversification method.

**Image fusion (IF)**. The radiological images in different chest CT scans present different modalities. The CoSinGAN, trained only on the single radiological image of modality 1, cannot smoothly generate radiological images of modality 2. Thus, we propose to fuse the synthesized images from two CoSinGANs that are trained separately on the single image of modality 1 and on the other single image of modality 2 directly. Given the same input condition, the two CoSinGANs are able to generate paired images of two different modalities that match each other pixel-by-pixel. Accordingly, we can simply fuse the two paired images as follows without losing image details and sharpness:

$$I_f = \zeta \times I_1 + (1-\zeta) \times I_2, \tag{11}$$

where $\zeta$ is the fusion coefficient. We introduce diversity by randomly setting the value of $\zeta$ from 0.0 to 1.0 in our experiments. As can be seen, IF can synthesize radiological images of intermediate modalities. Moreover, the synthesized COVID-19 infection of IF are more realistic and may have more clinical relevance than those of RC. Therefore, extending CoSinGAN by IR may have the potential to realize learning deep models for COVID-19 diagnosis from few representative radiological images.

### 3.3 Experiments on automated COVID-19 diagnosis

Most studies have adopted either or both of classification models and segmentation models to realize the automated COVID-19 diagnosis [8]. The classification models input a radiological image and output a binary scalar, where 1 indicates COVID-19 infection and 0 represents no COVID-19 infection. In comparison, the segmentation models input a radiological image and output a ternary mask that indicates where the lung and COVID-19 locate. Thus, we conduct our experiments on both of classification and segmentation models to explore the feasibility of learning COVID-19 diagnosis from a single radiological image.

### 3.3.1 Experiments based on classification models

**Baselines.** We train two baseline classification networks, i.e., ResNet18 and ResNet50 [30], on the training set of our materials for automated COVID-19 diagnosis. Deep residual learning networks are popular network architectures, where ResNet18 is a lightweight version whereas ResNet50 is heavyweight version. Using such two baselines can make the experiment results more convincing.

**Training sets**. Given all the conditions in the training set, we use the per-trained CoSinGAN with modality 1 to synthesize radiological images, called originally synthesized training samples (O-STS). We then impose RC as described in section 3.2.4 on CoSinGAN to get the first version of diversified training samples, called RC-STS. Similarly, we use the IF method to get the second version of diversified training samples, called IF-STS. We also build one training set with the same single image of modality 1 that is used to train CoSinGAN, called Sin-TS, and build another training set with the same two images of modality 1 and modality 2 that are used to train the two individual CoSinGANs, called Two-TS. For convenience, we call the original complete training set of our materials OC-TS. We finally train the baseline networks on OC-TS, Sin-TS, Two-TS, O-STS, RC-STS and IF-STS separately. It is worth noting that our CoSinGAN is able to synthesize infinite number of radiological images theoretically, but we only use the conditions in OC-TS to obtain the same number of radiological images as OC-TS. It means that each training set contains the same number of training samples, i.e., 2846 radiological images, which ensures that the parameters of models are updated the same number of times during each training process.

**Training details and evaluation metrics**. We train these baseline models with 10 epochs and mini-batch of 16 by using Adam optimizer with the parameters of $\beta_1 = 0.9$ and $\beta_2 = 0.999$. We use an initial learning rate of 0.0001 that is linearly decayed by 20% each epoch after 5 epochs. All the models are trained and evaluated with input channel of 3 and image size of $256 \times 256$. We use the weights pre-trained on ImageNet to initialize the parameters of our models. We perform the same strong augmentation (SA) that is used to train CoSinGAN on Sin-TS, Two-TS, O-STS, RC-STS and IF-STS, and perform the same weak augmentation (WA) on OC-TS. The binary cross-entropy (BCE) loss is adopted. All these trained models are finally evaluated and compared on the test set (674 real chest CT slices from 5 CT scans) of our materials by the common used metrics, i.e., sensitivity, specificity, and accuracy. After that, we repeat the same training and evaluation process with input image size of $512 \times 512$.

**Results and discussions**. We report our evaluation results in **Table 2** and **Table 3**. First, in the training process of Sin-TS and Two-TS, we observe that the training loss decreases rapidly to less than 0.01 in 1 or 2 epochs even with strong data augmentation, thus leading to poor classification results as shown in the second and third rows of **Table 2** and **3**. Next, the synthesized images achieve consistently better classification accuracy than Sin-TS and Two-TS. Specifically, RC-STS, synthesized by randomizing the input conditions of CoSinGAN, achieves slightly superior accuracy over the originally synthesized training samples O-STS except for the case of ResNet50 trained with image size of $512 \times 512$ (the last columns in the third and fourth rows of **Table 3**). We argue it is because that RC can synthesize radiological images with diverse appearance of COVID-19 infection, which facilitate the training of deep models for COVID-19 diagnosis; however, such synthesized COVID-19 infection, not confirmed in clinical, may mislead deep models. At last, IF-STS obtained by fusing the paired images of two different modalities from two CoSinGANs achieves notable classification accuracy of COVID-19 infection, significantly better than Sin-ST, Two-ST, O-STS and RC-STS, and even comparable to OC-TS. Such results confirm that our CoSinGAN have the potential to realize learning COVID-19 diagnosis from few representative radiological images.

**Table 2.** Evaluation results of the baseline classification networks trained on different training sets with image size of 256×256. The second column in this table represents the number of real samples that are used in the entire training process (including the training process of CoSinGANs). As can be seen, the fused radiological images synthesized by our CoSinGAN using only two real images achieve notable classification accuracy of COVID-19 infection. The 95% confidence intervals for evaluation results on 5 CT scans in the test set of our materials are calculated by using Student's t-distribution with (5 − 1) degrees of freedom (although values larger than 1.0 and smaller than 0.0 are meaningless, we can use them to highlight the differences between different results).

| Training set | The number of real images | ResNet18 | | | ResNet50 | | |
|---|---|---|---|---|---|---|---|
| | | Sensitivity | Specificity | Accuracy | Sensitivity | Specificity | Accuracy |
| Sin-TS | 1 | 1.000 (NaN, NaN) | 0.000 (NaN, NaN) | 0.608 (-0.095, 1.281) | 1.000 (NaN, NaN) | 0.000 (NaN, NaN) | 0.608 (-0.095, 1.281) |
| O-STS | 1 | 0.729 (-0.313, 1.538) | 0.420 (-0.224, 1.656) | 0.608 (0.044, 1.177) | 0.907 (0.531, 1.252) | 0.326 (-0.324, 1.608) | 0.680 (0.288, 1.201) |
| RC-STS | 1 | 0.685 (-0.528, 1.771) | 0.511 (-0.062, 1.569) | 0.617 (-0.144, 1.349) | 0.954 (0.863, 1.076) | 0.348 (-0.510, 0.913) | 0.717 (-0.037, 1.348) |
| Two-TS | 2 | 1.000 (NaN, NaN) | 0.000 (NaN, NaN) | 0.608 (-0.095, 1.281) | 1.000 (NaN, NaN) | 0.015 (-0.043, 0.076) | 0.614 (-0.071, 1.273) |
| IF-STS | 2 | 0.876 (0.552, 1.263) | 0.723 (-0.442, 1.474) | 0.816 (0.527, 1.084) | 0.959 (0.855, 1.088) | 0.337 (-0.478, 1.049) | 0.715 (0.279, 1.170) |
| OC-TS | 2846 | 0.980 (0.908, 1.050) | 0.576 (0.169, 1.090) | 0.822 (0.333, 1.254) | 0.961 (0.824, 1.090) | 0.792 (0.577, 1.072) | 0.895 (0.751, 1.005) |

**Table 3.** Evaluation results of the baseline classification networks trained on different training sets with image size of 512×512. The second column in this table represents the number of real samples that are used in the entire training process (including the training process of CoSinGANs). As can be seen, the fused radiological images synthesized by our CoSinGAN using only two real images achieve notable classification accuracy of COVID-19 infection. The 95% confidence intervals for evaluation results on 5 CT scans in the test set of our materials are calculated by using Student's t-distribution with (5 − 1) degrees of freedom (although values larger than 1.0 and smaller than 0.0 are meaningless, we can use them to highlight the differences between different results).

| Training set | The number of real images | ResNet18 | | | ResNet50 | | |
|---|---|---|---|---|---|---|---|
| | | Sensitivity | Specificity | Accuracy | Sensitivity | Specificity | Accuracy |
| Sin-TS | 1 | 1.000 (NaN, NaN) | 0.000 (NaN, NaN) | 0.608 (-0.095, 1.281) | 1.000 (NaN, NaN) | 0.000 (NaN, NaN) | 0.608 (-0.095, 1.281) |
| O-STS | 1 | 0.939 (0.798, 1.130) | 0.201 (-0.430, 0.704) | 0.650 (-0.060, 1.293) | 0.976 (0.908, 1.064) | 0.299 (-0.336, 0.775) | 0.711 (0.220, 1.163) |
| RC-STS | 1 | 0.998 (0.991, 1.006) | 0.258 (-0.401, 0.708) | 0.708 (-0.046, 1.349) | 0.237 (-0.419, 0.736) | 0.936 (0.846, 1.107) | 0.510 (-0.254, 1.210) |
| Two-TS | 2 | 1.000 (NaN, NaN) | 0.000 (NaN, NaN) | 0.608 (-0.095, 1.281) | 1.000 (NaN, NaN) | 0.000 (NaN, NaN) | 0.608 (-0.095, 1.281) |
| IF-STS | 2 | 0.778 (0.324, 1.224) | 0.780 (0.575, 1.139) | 0.779 (0.453, 1.158) | 0.863 (0.648, 1.147) | 0.538 (-0.252, 1.188) | 0.736 (0.358, 1.128) |
| OC-TS | 2846 | 0.854 (0.606, 1.147) | 0.947 (0.447, 1.317) | 0.890 (0.793, 1.012) | 0.922 (0.767, 1.075) | 0.917 (0.070, 1.484) | 0.920 (0.864, 0.979) |

### 3.3.2 Experiments based on segmentation models

**Baselines.** We train two baseline segmentation networks, i.e., ENet [29] and UNet [19], on the training set of our materials. ENet is a well-known segmentation network that has shown a good trade-off between accuracy and inference speed [29][31]. UNet is one of the most successful segmentation framework in medical imaging. In comparison, ENet is a lightweight network whereas UNet is a heavyweight network. Using such two baselines are much easier to obtain convincing results.

**Training sets.** We use the same training sets as detailed in section 3.3.1, including OC-TS, Sin-TS, Two-TS, O-STS, RC-STS and IF-STS. It is worth noting that each training set contains the same number of training samples, i.e., 2846 radiological images, which ensures that the parameters of models are updated the same number of times during each training process.

**Training details and evaluation metrics.** We train these baseline models with 50 epochs by using Adam optimizer with the parameters of $\beta_1 = 0.9$ and $\beta_2 = 0.999$. We adopt mini-batch of 8 for ENet and mini-batch of 2 for UNet respectively due to memory limitation. We use an initial learning rate of 0.0001 that is linearly decayed by 4% each epoch after 25 epochs. All the models are trained with input channel of 1 and image size of 256×256 from scratch. We perform the same strong augmentation (SA) that is used to train CoSinGAN on all training sets. Besides, the category-weighted cross entropy loss is adopted to emphasize the optimization of COVID-19 infection segmentation, where the weights of background, lung and COVID-19 infection are set to 0.1, 1.0 and 5.0. All these trained models are finally evaluated and compared on the test set (674 real chest CT slices from 5 CT scans) of our materials by Dice similarity coefficient (DSC). Meanwhile, we also compute the DSC scores on the subset of modality 1 and on the subset of modality 2 separately to make a more detailed comparison.

**Table 4**. Evaluation results of the baseline segmentation networks. The second column in this table represents the number of real samples that are used in the entire training process (including the training process of CoSinGANs). As can be seen, the fused radiological images synthesized by our CoSinGAN using only two real annotated images achieve notable segmentation accuracy of the lung and COVID-19 infection. The 95% confidence intervals for overall evaluation results on 5 CT scans in the test set of our materials are calculated by using Student's t-distribution with (5 – 1) degrees of freedom (although values larger than 100.0 and smaller than 0.0 are meaningless, we can use them to highlight the differences between different results).

| Training set | The number of real images | ENet | | | | | | UNet | | | | | |
|---|---|---|---|---|---|---|---|---|---|---|---|---|---|
| | | Modality 1 | | Modality 2 | | Overall | | Modality 1 | | Modality 2 | | Overall | |
| | | Lung | Infection | Lung | Infection | Lung | Infection | Lung | Infection | Lung | Infection | Lung | Infection |
| Sin-TS | 1 | 73.9 | 28.7 | 0.0 | 0.0 | 50.1 (-71.1, 130.9) | 19.4 (-29.6, 53.1) | 78.1 | 34.0 | 8.0 | 1.6 | 55.5 (-60.3, 132.3) | 23.5 (-34.9, 64.4) |
| O-STS | 1 | 81.5 | 49.4 | 7.3 | 0.9 | 57.5 (-65.7, 139.1) | 33.7 (-46.3, 86.6) | 72.5 | 59.5 | 1.0 | 0.6 | 49.4 (-68.7, 128.5) | 40.5 (-56.9, 105.2) |
| RC-STS | 1 | 75.5 | 50.7 | 6.7 | 4.3 | 53.2 (-63.7, 130.3) | 35.7 (-38.0, 86.2) | 73.9 | 57.9 | 6.5 | 25.2 | 52.0 (-56.7, 125.4) | 47.3 (-29.9, 121.5) |
| Two-TS | 2 | 70.1 | 24.5 | 74.1 | 39.8 | 71.4 (40.9, 103.4) | 29.4 (-4.7, 78.9) | 68.9 | 34.4 | 59.2 | 17.1 | 65.8 (32.5, 93.7) | 28.8 (-22.4, 67.9) |
| IF-STS | 2 | 75.3 | 40.3 | 81.0 | 36.2 | 77.1 (62.9, 94.5) | 39.0 (23.5, 52.6) | 76.1 | 57.8 | 74.0 | 48.4 | 75.4 (54.0, 94.3) | 54.8 (36.7, 69.1) |
| OC-TS | 2846 | 87.6 | 68.2 | 76.6 | 38.6 | 84.0 (54.7, 107.1) | 58.6 (-7.8, 119.7) | 90.8 | 77.3 | 87.8 | 75.2 | 89.5 (75.0, 103.0) | 76.6 (56.9, 99.4) |

**Results and discussions**. The segmentation scores measured by DSC are detailed in **Table 4**. As can be seen, the synthesized training sets, including O-STS, RC-STS, and IF-STS, consistently outperform Sin-TS and Two-TS by a large margin in COVID-19 infection segmentation. Considering the domain discrepancy between different modalities, we first compare Sin-TS, O-STS, and RC-STS that all use one real image of modality 1 specifically on the subset of modality 1; we find that O-STS and RC-STS achieve notable infection segmentation scores, much higher than (more than 20%) Sin-TS and even comparable (less than 20%) to OC-TS that contains 2846 real images. Such results implicate that the deep segmentation models trained on synthesized samples from CoSinGAN can generalize to the other image modalities better than the same models trained on a single real image directly by using strong data augmentation. Besides, we also notice that RC-STS obtains higher infection segmentation scores than O-STS, and such gaps are more obvious on the subset of modality 2 (3.4% for ENet and 24.6% for UNet). We argue it is caused by the using of proposed RC method (i.e., randomizing input condition of CoSinGAN) in RC-STS. We design the RC method to enable CoSinGAN to generate diverse samples, and expect to improve the generalization ability of deep models trained on synthesized samples. Thus, such results confirm the efficacy of RC. Next, we compare the results of Two-TS and IF-STS that both use an additional real image of modality 2. We observe that the additional real image significantly improve the infection segmentation scores on the subset of modality 2. Besides, we find that IF-STS achieve notable infection segmentation scores, much higher (9.6% for ENet and 26% for UNet) than Two-TS and even approximating (gap of 19.6% for ENet and gap of 21.8% for UNet) to OC-TS that contains 2846 real images. Such results strongly confirm that our methods have the potential to reduce the segmentation performance gap between deep models trained on extremely small image dataset and on large image dataset.

## Conclusion

The highly contagious COVID-19 has spread rapidly and overwhelmed healthcare systems across the world. Automated infection measurement and COVID-19 diagnosis at the early stage is critical to prevent the further evolving of COVID-19 pandemic. Unfortunately, collecting large training data systematically in the early stage is difficult. To address this problem, in this paper we explore the approaches of learning deep models for COVID-19 diagnosis from a single radiological image by resorting to synthesizing diverse radiological images. We propose CoSinGAN that can learn the conditional distribution of visual finds of COVID-19 infection from a single radiological image precisely and synthesize diverse, high-resolution and high-quality radiological images with COVID-19 infection effectively. Both deep classification and segmentation networks trained on synthesized samples from CoSinGAN (using 1 or 2 real images) achieve notable detection accuracy of COVID-19 infection. It strongly confirm that our method has the potential to realize learning deep models for COVID-19 diagnosis from few radiological images in the early stage of COVID-19 pandemic.

Due to the strong ability in learning conditional distribution of visual finds of COVID-19 infection from a single radiological image, our CoSinGAN can also be used to perform semantic manipulation, for instance, the addition and removal of COVID-19 infection. By adding COVID-19 infection to the off-the-shelf radiological images, we may obtain training samples that are much more diverse and thus may achieve much better detection accuracy of COVID-19 infection.

## Inference

[1]. J. T. Wu, K. Leung, and G. M. Leung, "Nowcasting and forecasting the potential domestic and international spread of the 2019-ncov outbreak originating in wuhan, china: a modelling study,"

The Lancet, vol. 395, no. 10225, pp. 689–697, 2020.
[2]. Z. Xu, L. Shi, Y. Wang, J. Zhang, L. Huang, C. Zhang, S. Liu, P. Zhao, H. Liu, L. Zhu et al., "Pathological findings of covid-19 associated with acute respiratory distress syndrome," The Lancet respiratory medicine, vol. 8, no. 4, pp. 420–422, 2020.
[3]. H. Shi, X. Han, N. Jiang, Y. Cao, O. Alwalid, J. Gu, Y. Fan, and C. Zheng, "Radiological findings from 81 patients with covid-19 pneumonia in wuhan, china: a descriptive study," The Lancet Infectious Diseases, vol. 20, no. 4, pp. 425 – 434, 2020. [Online]. Available: http://www.sciencedirect.com/science/article/pii/S1473309920300864.
[4]. X. Xie, Z. Zhong, W. Zhao, C. Zheng, F. Wang, and J. Liu, "Chest CT for typical 2019-ncov pneumonia: relationship to negative RT-PCR testing," Radiology, p. 200343, 2020.
[5]. T. Ai, Z. Yang, H. Hou, C. Zhan, C. Chen, W. Lv, et al., "Correlation of Chest CT and RT-PCR Testing in Coronavirus Disease 2019 (COVID-19) in China: A Report of 1014 Cases," Radiology, p. 200642, 2020.
[6]. Y. Li and L. Xia, "Coronavirus disease 2019 (covid-19): Role of chest ct in diagnosis and management," American Journal of Roentgenology, pp. 1–7, 2020.
[7]. Y. Fang, H. Zhang, J. Xie, M. Lin, L. Ying, P. Pang, and W. Ji,"Sensitivity of chest ct for covid-19: comparison to rt-pcr," Radiology, p. 200432, 2020.
[8]. Shi, Feng, et al. "Review of artificial intelligence techniques in imaging data acquisition, segmentation and diagnosis for covid-19." IEEE reviews in biomedical engineering (2020).
[9]. Mei, Xueyan, et al. "Artificial intelligence-enabled rapid diagnosis of COVID-19 patients." medRxiv (2020).
[10]. Oh, Yujin, Sangjoon Park, and Jong Chul Ye. "Deep learning covid-19 features on cxr using limited training data sets." IEEE Transactions on Medical Imaging (2020).
[11]. J. P. Kanne, "Chest CT findings in 2019 novel coronavirus (2019-nCoV) infections from Wuhan, China: key points for the radiologist," Radiology, p. 200241, 2020.
[12]. Kang, Hengyuan, et al. "Diagnosis of coronavirus disease 2019 (covid-19) with structured latent multi-view representation learning." IEEE Transactions on Medical Imaging (2020).
[13]. Yan, Qingsen, et al. "COVID-19 Chest CT Image Segmentation--A Deep Convolutional Neural Network Solution." arXiv preprint arXiv:2004.10987 (2020).
[14]. Ouyang, Xi, et al. "Dual-Sampling Attention Network for Diagnosis of COVID-19 from Community Acquired Pneumonia." arXiv preprint arXiv:2005.02690 (2020).
[15]. I. D. Apostolopoulos and T. A. Mpesiana, "COVID-19: automated detection from x-ray images utilizing transfer learning with convolutional neural networks," Physical and Engineering Sciences in Medicine, p. 1, 2020.
[16]. Loey, Mohamed, Florentin Smarandache, and Nour Eldeen M Khalifa. "Within the Lack of Chest COVID-19 X-ray Dataset: A Novel Detection Model Based on GAN and Deep Transfer Learning." Symmetry 12.4 (2020): 651.
[17]. Waheed, Abdul, et al. "CovidGAN: Data Augmentation Using Auxiliary Classifier GAN for Improved Covid-19 Detection." IEEE Access 8 (2020): 91916-91923.
[18]. Shaham, Tamar Rott, Tali Dekel, and Tomer Michaeli. "Singan: Learning a generative model from a single natural image." Proceedings of the IEEE International Conference on Computer Vision. 2019..
[19]. Olaf Ronneberger, Philipp Fischer, and Thomas Brox. U-net: Convolutional networks for biomedical image segmentation. In International Conference on Medical image computing and computer-assisted intervention, pages 234-241. Springer, 2015.
[20]. Isola P, Zhu J, Zhou T, et al. Image-to-Image Translation with Conditional Adversarial Networks[C]. computer vision and pattern recognition, 2017: 5967-5976.
[21]. Simonyan, Karen, and Andrew Zisserman. "Very deep convolutional networks for large-scale image recognition." arXiv preprint arXiv:1409.1556 (2014).
[22]. Z. Wang, E. P. Simoncelli, and A. C. Bovik, "Multiscale structural similarity for image quality assessment," in Proc. IEEE Asilomar Conf. Signals, Syst., Comput., vol. 2. Ieee, 2003, pp. 1398–1402.
[23]. H. Zhao, O. Gallo, I. Frosio, and J. Kautz, "Loss functions for image restoration with neural networks," IEEE Trans. Comput. Imaging, vol. 3, no. 1, pp. 47–57, 2017.
[24]. Z. Wang, A. C. Bovik, H. R. Sheikh, and E. P. Simoncelli, "Image quality assessment: from error visibility to structural similarity," IEEE Trans. Image Process., vol. 13, no. 4, pp. 600–612, 2004.
[25]. A. Dosovitskiy and T. Brox. Generating images with perceptual similarity metrics based on deep networks. In Advances in Neural Information Processing Systems (NIPS), 2016. 2, 3, 5.
[26]. J. Johnson, A. Alahi, and L. Fei-Fei. Perceptual losses for real-time style transfer and super-resolution. In European Conference on Computer Vision (ECCV), 2016. 2, 3, 4, 5, 8, 14.
[27]. Wang, Ting-Chun, et al. "High-resolution image synthesis and semantic manipulation with conditional gans." Proceedings of the IEEE conference on computer vision and pattern recognition. 2018.
[28]. Ma, Jun, et al. "Towards Efficient COVID-19 CT Annotation: A Benchmark for Lung and Infection Segmentation." arXiv preprint arXiv:2004.12537 (2020).
[29]. Paszke, A., Chaurasia, A., Kim, S., Culurciello, E., 2016. Enet: a deep neural network architecture for real-time semantic segmentation. arXiv: 1606.02147.
[30]. He K , Zhang X , Ren S , et al. Deep Residual Learning for Image Recognition[C]// IEEE Conference on Computer Vision & Pattern Recognition. IEEE Computer Society, 2016.

[31]. Kervadec, Hoel, et al. Constrained-CNN losses for weakly supervised segmentation. Medical image analysis 54 (2019): 88-99.